**Spatio-temporal evolution of electric field inside a microwave discharge plasma during initial phase of ignition and its effect on power coupling**

Chinmoy Mallick[1,2,a], Mainak Bandyopadhyay[1,2,a], and Rajesh Kumar[1,2,a]
[1]*Institute for Plasma Research (IPR), Gandhinagar, Gujarat – 382 428, India*
[2]*Homi Bhabha National Institute (HBNI), Mumbai, Maharastra, India*

[a)] Electronic mail: chinmoyju1990@gmail.com; mainak@ipr.res.in; rkumar@ipr.res.in

**Abstract**

During the initial phase of microwave (MW) power launch inside a MW discharge ion source (MDIS), plasma and the electric fields are evolved with time together in the plasma volume. The spatio-temporal evolution pattern in the cavity of a MDIS is reported here, highlighting the role of these electric fields on power coupling processes. Evolution of electric field and so power coupling processes are calculated using Finite Element Method (FEM). Unlike PIC/MCC or hybrid fluid, here FEM model uses time dependent Poisson solver through drift-diffusion approach. In presence of plasma gradient in the volume ambipolar electric field is generated which interacts with the MW electric field to form resultant total electric field which continuously vary during this evolution period. It is observed that main power coupling mechanism is electron cyclotron resonance (ECR) method, however with the evolution of plasma, the mode shifts from ECR to off-ECR type heating with time. Off-ECR heating in the form of upper-hybrid resonance (UHR) method, electro-static (ES) ion acoustic wave heating method are important heating mechanisms during the over-dense plasma condition, when density is above critical density for launched MW frequency, 2.45 GHz. An indirect verification of simulated temporal evolution of hot electron temperature and plasma density with the experiment, is done under similar configuration and operating environment. The experimental results are found to be agreed reasonably well with the simulation in the low power range.

## I. INTRODUCTION

Microwave ECR plasma source is a very popular plasma reactor for industrial as well as research applications, particularly in the accelerator field [1-3]. MW ion source operated in pulsed mode has few branches of applications, such as production of rare radioactive ion beam [2] and intense heavy ion beams [3]. The basic mechanism to ignite and sustain the plasma in a high frequency discharge is the power absorption by the electrons from the space and time dependent electric field. In pulsed MW plasma, spatio-temporal evolution of electric field inside the plasma chamber and so the power coupling process to characterize the plasma dynamics is important to optimize the performance of the pulsed MW discharge device used for ion source. Published experimental evidences [2] prove that a transient peak of extracted ion current occurs at the beginning of the MW pulse (pre-glow mode). On the other hand the intense heavy ion beam which is the requirement of Large Hadron Collider (LHC) experiments [3] at CERN, is extracted from ECR ion source during the time-interval when the MW pulse is turned off (after-glow mode). During the transient incident when MW is just launched, and the plasma is evolving, the MW launcher observed transient impedance load which affects it's power coupling efficiency. Cortazer et.al. had presented [4] experimental evidences qualitatively on the time evolution of the average electric field within few µs, when the plasma density is increasing. They showed that an average electric field inside plasma loaded cavity is dropped by ~ 60% of the unloaded value during plasma formation time t = ~20 µs [5].

Despite of having some research work in last few decades [6-10], a detailed picture on space-time evolution of each component of the electric fields affecting different power transfer mechanisms during the gas ignition moment ( ns to µs) is still missing in the low pressure regime. Although the kinetic models like PIC/MCC or even the hybrid fluid/PIC gives more precise and efficient results in the MW plasma discharge, it demands intensive computational hardware because of its particle approach. These models are inappropriate in estimating fast electron dynamics at lower pressure. Given the limitations of fluid modeling approach [11], here FEM based model used can give efficient results at lower pressure using Time Dependent, Partial Differential Equation solver (TDPDE). Similar approach is used by G. J. M. Hagelaar et.al. [12] but for steady state case to study 2D space dependent power absorption mechanisms when plasma density is raised beyond cut-off density for the MW.

The main objectives in this study is to investigate the evolution of the electric field and its role on the hot electron population build-up. The spatio-temporal profile of the electric field may provide information about the associated different MW-plasma coupling mechanisms during gas ignition periods. The different MW-plasma coupling mechanisms (ECR, UHR and electric field polarity reversal associated with ES wave heating) during the plasma density evolution after the MW launch (t = 0 s) can be understood from the behaviors of electric fields. Also, in order to be consistent with the simulated results, the hot electron and plasma density variation during the evolution processes is verified with the experiment in an indirect way.

The manuscript is arranged in the following manner. Section – II deals with the description of the simulation tools, domain and the procedure. Section-III describes the experimental methods used to verify the simulated temporal plasma parameters. Section – IV discusses the simulated results in terms of spatio-temporal evolution of plasma, different components of total electric field present inside the plasma and corresponding power coupling processes which is succeeded by the verification of the simulated results with experiments. Finally summary and conclusions are drawn in section–V.

## II. Simulation modelling of MW-plasma interaction:

Figure 1 shows complete computational domain of the MW-plasma interaction in a microwave discharge ion source. Computational geometry, magnetic system and microwave frequency are kept almost similar to the experimental configuration described in [13]. Here, finite element based COMSOL MW-plasma model is very efficient unlike PIC/MCC or hybrid fluid model which demands intensive computational hardware and time consuming runs. During modelling, the MW propagation and plasma parameters evolution are considered to be decoupled to each other in the calculation because of timescales. The oscillating MW electric fields ($\tilde{E}$) are averaged out after some MW periods neglecting the plasma evolution. The resultant electric fields which become periodic are taken as input to the MW plasma model. The time integration in the electron's momentum equation is continued along with the Maxwell's equations over a few MW periods until the solution becomes periodic so that an average power can be transferred to the electrons over such a period. A detailed description about the power absorption through the MW electric field is given in [12] in which an average absorbed power, which is the input to the plasma model, updates their spatial profile periodically during simulation. This process is continued until a steady state solution is achieved for at least ~$10^3$ ($\omega/\nu_m \sim 10^3$) MW periods. Here, ω is MW frequency and $\nu_m$ is electron's momentum transfer frequency. The electrons pass through the resonance zone in a very short time due to its thermal motion which results in non-local kinetics effects and de-phasing between the velocity and field oscillations.

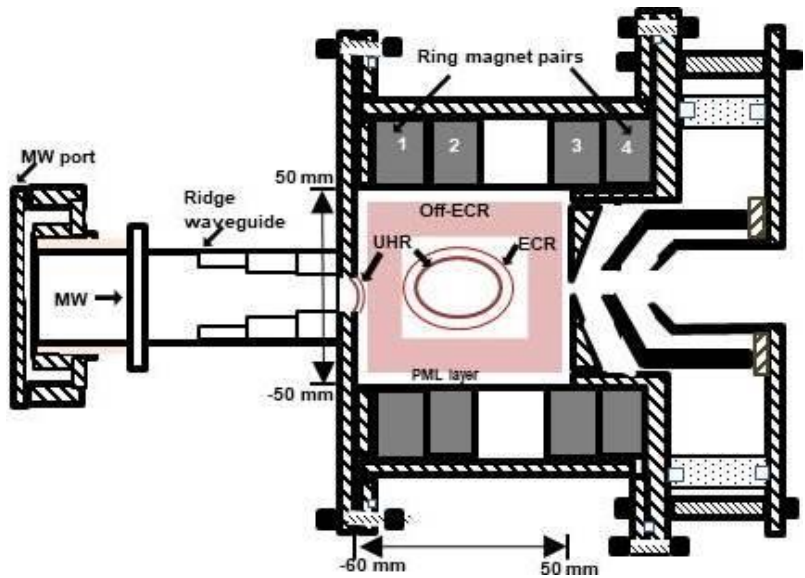

Figure 1: Computational domain of MW discharge ion source

This is difficult to describe in fluid model. To overcome this problem, the effective collision frequency ($\nu_{eff}$) is introduced to get a convergence in the simulation[20]. In order to be consistent with the collision less heating, $\nu_{eff}$ must be of the order of inverse transit time through the resonance region. This $\nu_{eff}$ depends on the thermal speed and the magnetic field gradient whose values are close to operating frequency considering the experimental parameters.

In the present MW plasma model, heating through MW $\tilde{E}$ field is considered as the same way as described in [12]. The MW equations are solved in frequency domain and all other parameters in time domain. We have started from Maxwell's equations to justify this modeling approach. All the variable fields are varying with time having frequency[20] $\omega/2\pi$. Here, the total electric field (E) (in µs time zone) which is the superposition of MW and ambipolar electric field gets modified with time by the evolution of the plasma while propagating through it and is solved following the equations given below [14].

$$\nabla \times \mu^{-1}(\nabla \times E) - k_0^2 \left(\varepsilon_r - \frac{i\sigma}{\omega\varepsilon_0}\right) E = 0 \ . \tag{1}$$

Here $k_0$ is vacuum wavenumber, $\varepsilon_r$ is relative permittivity, E is total electric field (V/m) and σ is full tensor plasma conductivity. Also σ is a function of plasma density, collision frequency and magnetic flux density. Hence, if the plasma density, collision frequency, magnetic field components are known at a particular plasma loading during the plasma evolution, all the different components of the E-field can be determined. Since the MW $\tilde{E}$ field modification during the sub-nanosecond region is separated from the electric field during plasma evolution, the presented electric fields which are in the microsecond region consists of both MW as well as ambipolar type. Perfectly matched layer condition is used for microwave propagation to represent infinite space. In this model, electron transport properties follow the Boltzmann distribution which is an integro-differential equation in phase space (r, u), can't be solved efficiently. For solving it, FEM method considering plasma as a fluid approach with the drift diffusion approximation, is adapted from Boltzmann equation by multiplying some weighing function and then integrating it over the velocity space. This results in completely three-dimensional and time dependent equations. While modeling, ion motion is assumed to be negligible because of large mass with respect to the electron motion in MW timescale (ns). Electron density is also assumed constant in space in resonance zone and Debye length is much smaller than the MW interaction length. The mean electron velocity in MW time scale is obtained by assuming Maxwellian Distribution function and taking first derivative of the Boltzmann Equation as given in [14-15].

The following drift-diffusion equations are used to compute the electron density (or plasma density) and electron energy density.

$$\frac{\partial n_e}{\partial t}+\nabla\cdot[-n_e(\mu_e\cdot E)-D_e\cdot\nabla n_e]=R_e \tag{2}$$

$$\frac{\partial n_\varepsilon}{\partial t}+\nabla\cdot[-n_\varepsilon(\mu_\varepsilon\cdot E)-D_\varepsilon\cdot\nabla n_\varepsilon]+E\cdot\Gamma_e=R_\varepsilon \tag{3}$$

Where $E\cdot\Gamma_e=en_ev_e\cdot E_{ambipolar}-\Pi$ [14]. The heating term $E\cdot\Gamma_e$ in electron motion in equation (3) has two components. First part gives energy to electrons by means of the ambipolar fields. Second part ($\Pi$) is absorbed power density ($n_e\langle v_e\cdot\tilde{E}\rangle$) of the electrons by the MWs. Here, $n_e$ is electron density (m$^{-3}$), $v_e$ is mean electron velocity (m.s$^{-1}$). $R_e$ is the source term, ionization rate (m$^{-3}$.s$^{-1}$), $\Gamma_e$ is the electron flux (m$^{-2}$.s$^{-1}$), $\mu_e$ is the electron mobility (m$^2$.V$^{-1}$.s$^{-1}$), $\mu_\varepsilon$ is electron energy mobility, E is the total electric field, including MW and ambipolar electric field (V.m$^{-1}$), $D_e$ is the electron diffusivity (m$^2$.s$^{-1}$), $n_\varepsilon$ denotes the electron energy density (V.m$^{-3}$), $R_\varepsilon$ is the energy loss/gain due to inelastic collisions (V.m$^{-3}$.s$^{-1}$), $\Gamma_\varepsilon$ is the electron energy flux (m$^{-2}$.s$^{-1}$), $n_\varepsilon$ is the electron energy mobility (m$^2$.V$^{-1}$.s$^{-1}$), and $D_\varepsilon$ is the electron energy diffusivity (m$^2$.s$^{-1}$).

Here, the electron diffusivity, electron mobility and its energy diffusivity is estimated following the relations written below.

$$D_e=\mu_eT_e; \quad \mu_\varepsilon=(5/3)\mu_e \quad \text{and} \quad D_\varepsilon=\mu_\varepsilon T_e \tag{4}$$

The electron transport properties are full tensor quantities. The full tensor term, electron mobility is a function of dc electron mobility (without magnetic field) and magnetic flux density. DC electron mobility ($\equiv 1\times 10^{25}/N_n$) and applied magnetic flux density are known quantities. Here, $N_n$ is neutral number density. The electron source term ($R_e$) (see equation 2) is obtained from plasma chemistry and is expressed below in terms of the rate coefficients as, $R_e=\sum_{j=1}^{M}x_jk_jN_nn_e$. Here, M, $x_j$ and $k_j$ are number of reactions, mole fraction of a particular species for reaction j and rate coefficient for reaction j respectively. The details of all reactions involved in the process is given in [14]. The energy loss term (see equation 3) for all reactions are given below as, $R_\varepsilon=\sum_{j=1}^{M}x_jk_jN_nn_e\Delta\varepsilon_j$, where $\Delta\varepsilon_j$ is energy loss from reaction j. These energy loss and electron source terms are inherently calculated by the COMSOL multiphysics interface. Whereas, the rate coefficients are obtained from the cross section data following the relation, $k_k=\gamma\int_0^\infty \varepsilon\,\sigma_k(\varepsilon)f(\varepsilon)d\varepsilon$; where, $\gamma$ is $(2q/m_e)^{1/2}$. Here, $m_e,\varepsilon,\sigma_k$ and f are electron mass and energy, reaction cross section and Maxwellian distribution function of electron energy respectively.

The plasma chamber wall is grounded. The reflections as well as the secondary emission and thermal emission of the electrons are neglected at the wall boundaries. As a result, the electron flux and electron energy flux at wall boundary are written as, $n\cdot\Gamma_e=(\frac{1}{2}v_{e,th}n_e)$ and $n\cdot\Gamma_\varepsilon=(\frac{5}{6}v_{e,th}n_e)$ respectively. Also, losses of heavy plasma particles at the walls and also their migration are assumed only due to the surface reactions and ambipolar electric field respectively. The plasma absorbed power ($P_{absorbed}$) is kept fixed by re-adjusting the normalization factor (α) at any moment during the plasma evolution ( ns to μs ) process, following the relation, $P_{absorbed} = \alpha \iiint n_e P_{set}\, dV$. Here, $P_{set}$ ($\equiv -e\langle \tilde{v}_e . \tilde{E}\rangle_t$) is the average set power. This normalization is done because of the convergence of the self-consistent solution to avoid a situation of plasma disproportionate power absorption. Here, fixed absorbed power of 70 W is chosen to compare and benchmark with the experimental findings reported in [13] under similar system and operating conditions.

In the calculation, refrence is taken at time t = 0 when MW is launched into the vacuum chamber whose magnetic field profiles are calculated by fixed solver available within COMSOL and used [13] in plasma model. MW is launched in ordinary mode through the ridge waveguide, the electric field is maximum at center and its propagation vector is assumed to be parallel to the static applied B-field. After launching immediately, MW starts interacting with the background gas in the range of their time periods (ns). To understand and visualize the modification of electrical field and other plasma parameters in the plasma volume with time, the total plasma breakdown incident is divided into some discrete time values, which are taken as, 10ns, ~67ns, ~158 ns, 452 ns, 630 ns, ~2 μs, 2.5 μs, 3 μs, ~5 μs, ~8 μs, ~20 μs, ~40 μs, 250 μs and 300 μs, from the very start of MW interaction with argon gas. The initial assumed plasma boundary conditions are: plasma density as $1\times10^{12} m^{-3}$ and plasma temperature as ~ 4eV for estimating plasma conductivity σ [21]. Equation (1) is used to calculate different components of the total electric field (E) at first instance based on these initial conditions, which later utilized to estimate the temporal variation of different plasma parameters and so elctric field components self-consistently.

Solvers are used sequentially to compute first the magnetic field and then the microwave plasma parameters. A flowchart of the computation is depicted in fig.2. The number of degrees of freedom used for solving the magnetic field is 54525. Equation based mesh adaption technique is employed to create extremely fine mesh size on the resonant surfaces. The maximum and minimum mesh element size for the magnetic field and microwave-plasma model are 2.4 mm and 0.2 mm respectively. The maximum and minimum mesh element size on different edges of the computational domain are 0.5 mm and 0.0055 mm respectively. The first computed magnetic field is used in microwave plasma model for studying the plasma tensor properties. The number of degrees of freedom solved for in the time dependent solver of the microwave plasma model is 47005.

During the calculation the concept of effective collision frequency $\nu_{eff}$ is used to describe the sudden phase-decoherence when phase relationship between velocity and the electric field oscillation is destroyed on the time scale of electrons experiencing the field variations. In the time scale of resonant cyclotron motion, the accelerating field acts on the electrons only for a very small time duration that they spend in the resonance region. In that zone the electrons also experience a spatial density variations which is responsible for radial ambipolar electric field originating on the resonance surface and so there is a possibility a de-phasing from the resonance condition. De-phasing may happen between electron gyro motion and the MW oscillatory electric field, resulting in deceleration of electrons when out of phase. This leads to temporal asymmetry of acceleration and decceleration in opposite cycle and reponsible for effective heating of the electrons. Effective collision frequency $\nu_{eff}$ can be estimated as, $\nu_{eff} = \sqrt{v_e \omega / \delta_B}$; where $v_e$ is the electron thermal speed, $\omega$ is the MW frequency and $\delta_B$ is the gradient of magnetic field. The value of $\nu_{eff}$ is few orders more than that of actual collisional frequency an electron suffers with background neutral gas molecules. However, it turns out that $\nu_{eff}$ has insignificant effect on power absorption profile but helps to overcome numerical instability. To obtain a steady-state solution over the number of MW periods required is $\omega/\nu_{eff}$, which computationally less rigorous compared to the case actual collisions, if considered.

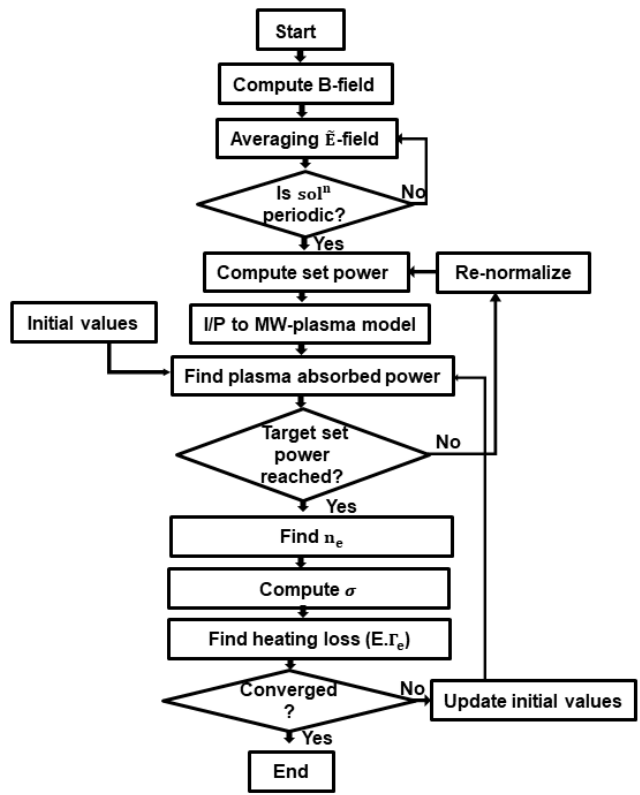

Figure 2: Simulation flowchart of MW plasma interaction in COMSOL Multiphysics

The radial ($E_r$) and axial ($E_z$) electric fields of MW, plays important role in the physics of collisionless heating of the plasma. These electric field components, which are parallel and perpendicular to the permanent magnetic field exhibits sharp rise and fall in the ECR surface. The sharp increase and decrease in electric field profile helps in accelerating and decellerating the electrons in the resonance region. To understand the resonance zone in the plasma chamber, magnetic field contours are shown in fig. 3.

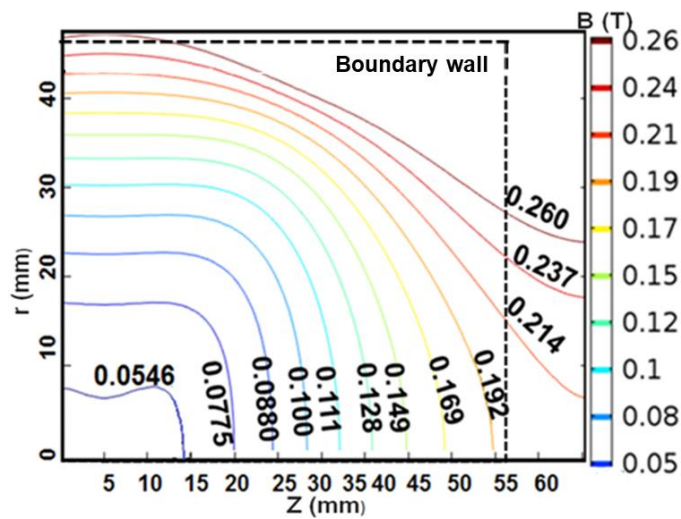

*Figure 3: Magnetic field contour inside the quadrant section of the plasma chamber which has cylindrical axis-symmetry. The B-field is also simulated using COMSOL software. The narrow contour area near 0.088T line is the ECR zone (corresponding to 0.0875T). The span of ECR zone is around z = ±24mm and r = ±28 mm.*

### III. Experimental method:

To verify the temporal behavior of simulated plasma density and hot electron temperature, experiments are performed under similar system, magnetic field and microwave configuration. A detailed description of the experimental setup and diagnostics methods are discussed in ref [13]. Real time measurement of plasma parameters in the *ns*-time scale demands some diagnostic hardware having fast time response (*ns* range) which are very advanced and so expensive. Here, in this experimental setup, an indirect verification of the simulated temporal data with the experiment is performed to check the consistency of the simulation results. At first, set power response of the 2.45 GHz magnetron in the frequency domain is measured at the directional coupler port [13] by a microwave Spectrum Analyzer (SA). The pulse response of the set power (e.g. 200 W) along with its rise time is obtained by doing inverse Fourier transform of the spectrum analyzer data. The hot electron temperature and plasma density is obtained from Langmuir probe diagnostic in the low set power level (50-200 W). The rise time of a particular set power (i.e., 50 W, 70 W, 130 W and 200 W) is confirmed from the plot of detected magnetron's set power vs. time. After that, the estimated hot electron temperature and density at different set powers 50 W, 70 W, 130 W and 200 W with their corresponding rise times *600ns, 720ns, 1230ns and 2200ns* are compared with the temporal results, obtained from simulation in section IV below.

## IV. Results and discussion

The temporal behaviour of the simulated total electric fields, total power deposition and corresponding temporal variation of plasma density upto the steady state condition are presented.

### A. Behaviors of total electric field ($|E|$) during plasma evolution:

Figure 4(a) and (b) represent radial profile of total electric field on different axial location at 3μs and 20μs instances after MW launch respectively. In figure 4(a), the radial profile at z = -40 mm, near the MW launching port (z = -60 mm in fig. 1) is maximum on the axis but reduces as we move towards the extraction zone z = 40mm due to plasma shielding as per available plasma density. Figure 4(a) also shows a sharp change in E-field at time t ~3μs near the 2.45 GHz ECR locations (r ≈ 23 mm) for z = ± 20 mm. Sharp gradient of E-field indicates MW power absorption near ECR zone [16]. The location of power absorption depends on the magnetic field profile within the chamber volume.

In figure 4(b), the radial profiles of total E-field for different axial location at 20μs are shown. The shape of the radial profile at z = -40 mm is similar to that of fig. 4(a) but with reduced value due to plasma shielding. Characteristic E-field gradient for z = ± 20 mm is shifted towards the off-ECR regime (r ≈ 28 mm). This indicates that power absorption location is shifted from ECR zone to off-ECR zone. The power absorption location shift from ECR zone to off-ECR zone or UHR zone with time is clearly visible in figures 5, where snap shots of power absorption contour within the plasma cavity volume at different time instances after MW launch are presented. During that time (t = 20 μs) plasma approaches towards steady state. The evolution of plasma density and temperature is shown in figure 6.

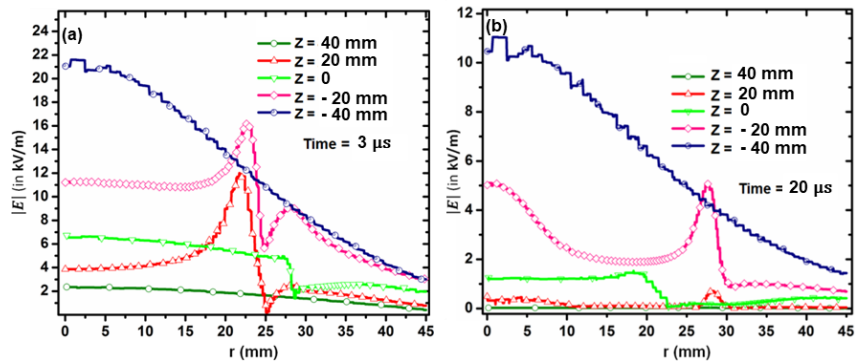

*Figure 4: (a) Total electric field ($|$E$|$) magnitude during gas ignition time of ~3μs, (b) Total electric field ($|$E$|$) magnitude during gas ignition time of ~20 μs.*

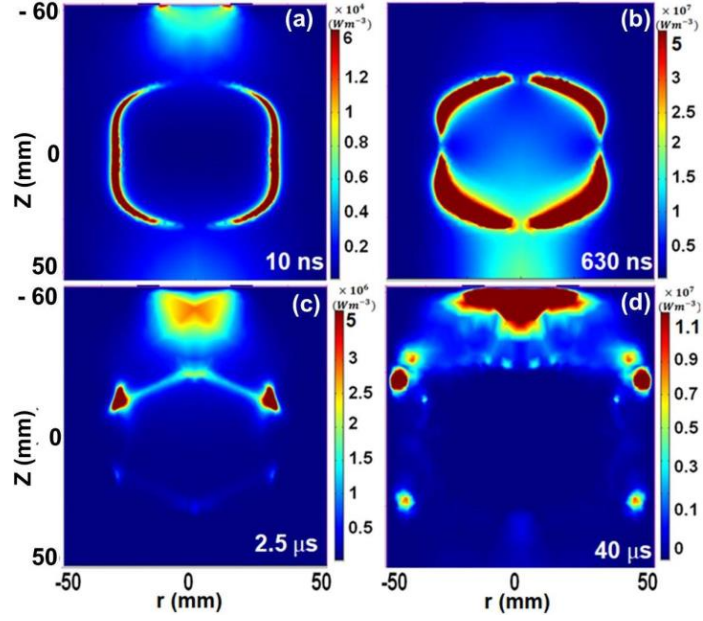

*Figure 5: (Color online) Power deposition density at different time steps for 70W of absorbed power. (a) t = 10 ns, peak power density is $6.6 \times 10^4$ $W/m^3$, (b) t = 630 ns, peak power density is $1.43 \times 10^7$ $W/m^3$, (c) t = 2.5 μs, peak power density is $3.12 \times 10^6$ $W/m^3$ and (d) t= ~40 μs, peak power density is $1.67 \times 10^7$ $W/m^3$.*

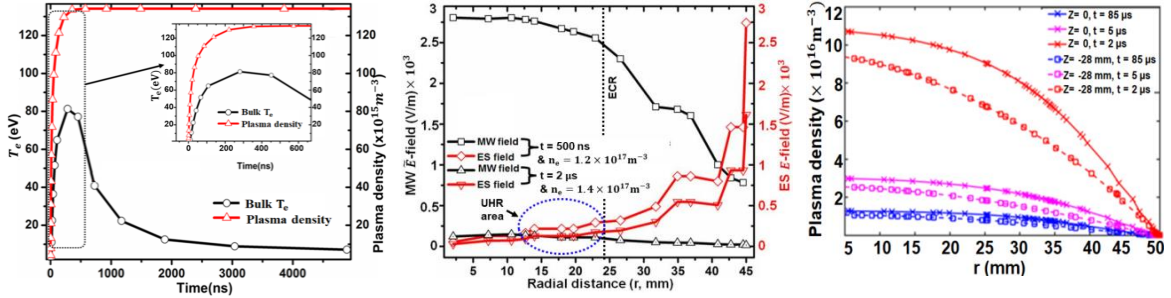

*Figure 6. (a) Temporal evolution of electron density and temperature during plasma formation time at point (r = 0, z = -28 mm) with gas pressure, $2 \times 10^{-3}$ mbar. (b) Temporal variation of MW electric field and plasma generated ambipolar electric field, the two main constituents of total electric field. (c) Radial profile of plasma density at three different time instants (t = ~2 μs, ~5 μs and 85 μs), at two axial locations during the evolution of plasma.*

### B. Time evolution of plasma with power deposition

In the beginning (figure 5a), when the plasma density is low, MW power is deposited everywhere on the 2.45 GHz ECR (~ 0.0875T) surface. With the elapse of time (see figure 5b-d), plasma density rises and correspondingly MW power deposition location gets shifted towards the off-ECR site or UHR zone (where the conditions $n_e < n_c$ and $B < B_{ECR}$ are satisfied [17]. The term $n_c$ stands for critical density which is $7.4 \times 10^{16} m^{-3}$ for 2.45GHz MW launched frequency and $B_{ECR}$ is 0.0875T. Comparing figure 5(c) and figure 6 (a), one notices that ~2.5 μs onwards the plasma density has become far more (~$1.3\times10^{17}$ m$^{-3}$) than the critical value ($7.4 \times 10^{16}$m$^{-3}$) and overdense condition is reached. This indicates that MW imparts very less amount of energy to the electrons on the ECR surface. Figure 6 (a) evidences that until ~630 ns (i.e. 0.63 μs) comes, the hot electron temperature (~75 eV) remains close to the maximum value (~80 eV), after that it falls sharply. Hence, the heating through ECR process continues to energize the electrons during this time. As a result, figure 5(b) shows highest absorbed power density (~$5\times10^7$ Wm$^{-3}$) on the ECR surface during time 630 ns. Correspondingly plasma density at the same instant of time is just above the critical density. The sharp decrease of hot electron temperature along with increase in the plasma density with time even above critical density) can be attributed to the off-ECR or ES surface wave heating in the power coupling process [18]. Fig.6(b) shows the temporal variation of MW electric field and plasma generated ambipolar electric field which is electro-static (ES) in nature during plasma evolution. ES electric field are responsible for UHR heating at the locations where magnetic field and plasma density meet the above specified conditions. To understand the role of different power coupling mechanisms (ECR, UHR, polarity reversal related to ES heating) during the plasma evolution, the total electric field is splitted up into different components ($E_r$ and $E_z$) and the spatio-temporal evolution of these different components are shown in figures 7 and 8.

Figure 6(c) shows variations of plasma density in radial direction on two axial locations for three different time instants, t = ~2, ~5 and ~85μs during the plasma evolution process. One location z = -28 mm is towards the MW launching port. The plasma density is approaching steady state, but plasma temperature is not yet stabilized during these time instances, as shown in fig. 6(a). It becomes clear that the plasma density is always higher at the axial position z = 0 mm than at the z = -28 mm. Theory as well as experiment proves that higher plasma density needs more energy absorption. This evidence is reflected in the radial as well as axial electric fields as shown in figures 7 and 8.

## C.   $E_r$-field evolution with time:

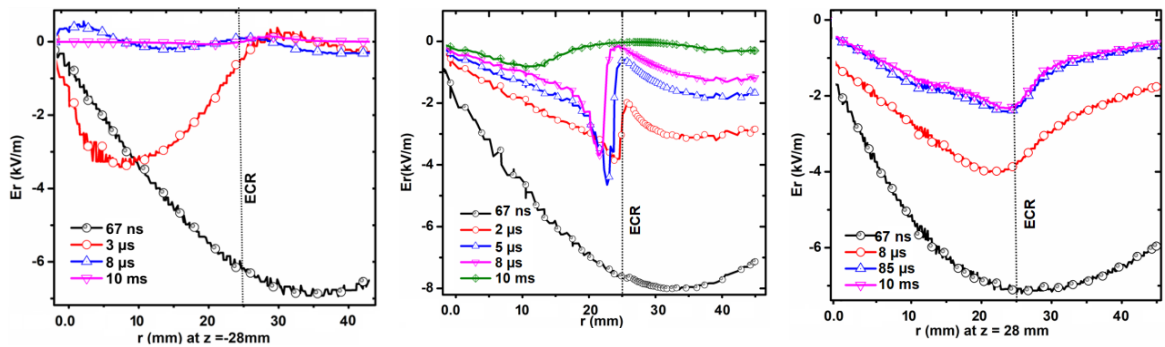

Figure 7: $E_r$-field evolution and its pattern during plasma generation (a) r = - 28 mm, (b) r = 0 mm and (c) r = 28 mm.

Figure 7 (a – c) shows the radial $E_r$ field profile on different "z" location on the axis of the MW discharge volume at different time instances after MW launch. Fig 7(a) profiles are at z = - 28 mm, closer to the MW launch port (z = -60 mm). Fig 7(b) profiles are at z = 0 mm, at the center of the plasma volume and fig 7(c) profiles are at z = 28 mm, closer to the beam extraction port. In all the figures, sharp change of $E_r$ field indicates the locations where power absorption takes place.

At t = 67 ns after starting of MW launch, the plasma density is below critical density and MW radial electric field is available almost through out the volume with a radial profile as per the system configuration (launching port and the magnetic field profile), as seen in all 7(a – c) figures. With advancement of time (> 500 ns) density crosses critical density and radial profile gets modified as per the density evolution pattern. Sharp change in electric field appears in those locations when power absorption conditions are favourable. A variation in radial locations in different figures are due to the variation of resonance magnetic field contour (see fig.3). Initially, the electrons are resonantly heated up by ECR phenomenon till time t = ~ 2μs. After that, this gradient gets shifted in the upper hybrid resonance ($f_{UHR} = \sqrt{(f_{pe}^2 + f_{ce}^2)}$) locations (n < $n_c$ and B < $B_{ECR}$), where launch frequency matches with $f^{UHR}$. Present simulated results shows plasma density after ~2 μs reaches overdense state at r = 0 and z =-24 location. But in some places where B < $B_{ECR}$, plasma density still remains below the critical density near this time (2μs). This is confirmed from the non-zero and high value of $E_r$-field near ECR zone during these time (~2 μs) as shown in figure 7(a). Hence, the double conditions i.e., B < $B_{ECR}$ and $n_e$ < $n_{critical}$ are met for the occurrence of UHR phenomenon. Figure 6(a) and (b) show that beyond 500 ns the ES component of the total E-field is due to the ambipolar field, grows with increase of plasma density and correspondingly the MW EM component decreases. With time power transfer is shifted from EM compoenet to ES component, which does not suffer and critical density criteria and so penetrate into the plasma volume even in overdense condition. The mode change of electric field nature from EM to ES is a signature of UHR heating [17]. Present simulated results shows the $E_r$-field gradient shifts towards the UHR (see figure 4 and 7) region after ~ 2 μs, generates hot electrons in the UHR zone (subject to the fulfillment of transition electric field and magnetic field gradient condition as described in [19]) which consequenctly interact with the slow extraordinary wave producing the cyclotron instability in plasma [2]. As the plasma density at that time instance reaches overdense state, a part of the launched ordinary wave after crossing the evanescent layer convert into the slow extraordinary wave in the UHR locations.

At t = 67 ns, the peak $E_r$ field amplitude at the ECR radial location is >7 kV/m at different z - locations. This value is enough for ionization of the Argon gas. At t = ~2μs, the $E_r$ field amplitude varies from 3.75 kV/m to 0.15 kV/m within ECR surface. This nonlinear variation of $E_r$ field on different resonant surfaces which transfer energy to plasma electrons agrees well with published reports given by J. Hopwood et. al.[6]. Power absorption phenomena in the plasma are also visible in the beam performance in terms of spike or instabilities. Ropponen et. al. presented that [3] the sharp transient in the ion current density during the preglow mode is accompanied by the sharp fall in the average hot electron temperature. Mansfeld [2] proved experimentally that the beam current oscillations due to cyclotron instability in the afterglow mode occur because of the interactions of hot electrons with the slow extraordinary wave produced in the UHR region.

In fig. 7(b) the sharp gradient part of the radial electric field is shifting towards central location (UHR region) with a velocity of 1250 m/s -1500 m/s which lies in the range of ion sound speed. Previous published reports say the electric field which propagates in the perpendicular direction with respect to the DC B-field direction, encounters UHR layer when the plasma is overdense. Here in this set up, for instance, the electric field in ordinary mode (O-mode) may get converted into slow extraordinary mode (X-mode) in front of the UHR layer through 'O – slow X' conversion process. This slow X mode can produce ion wave at the UHR layer [17].

### D.  $E_z$-field evolution in plasma with time:

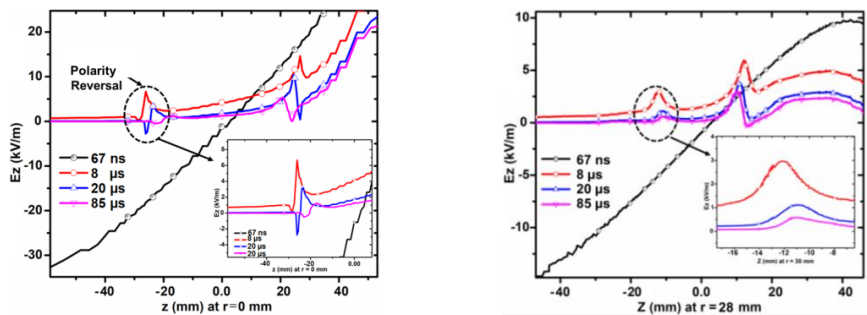

Figure 8 (Color online) The spatio-temporal evolution of $E_z$-field on the axis in the plasma volume. Two locations of sharp $E_z$ change indicate two power absorbing zones on the path (see fig.5 for visualization). (a) r =0 mm, (b) r= 28 mm.

As the MW is launched in ordinary mode through the ridge waveguide, the electric field is maximum at center and its propagation vector is parallel to the static applied B-field. The z-component of the total electric field at different time intstants of the MW plasma evolution processes are plotted in figure 8 (a) and (b) for different radial locations, r = 0 and 28 mm respectively. Figure 8(a) and (b) show a portion of one wavelength of $E_z$-field is distributed througout the plasma chamber during 67 ns and is having significant values. It is clear that $E_z$ field disappears inside the chamber as soon as the plasma starts creating during the plasma evolution. $E_z$ field probably doesn't have any significant contribution in accelerating the electrons, otherwise axial electron losses will occur. $E_z$ field after ~8μs is almost zero everywhere except near the MW launching side, till the nearest ECR surface, as shown in figure 8 (a). The decrease in $E_z$ field is more where the plasma density is high.

The polarity reversal near the ECR region of z-component of the total electric field near time t = ~ 20 μs is caused by the ambipolar field due to the density gradient which is computed form the electron momentum balance equation using drift-diffusion approach [20].The shifting of peaks is inward, toward the center in either of the power absorbing locations with time as shown in figs 8(a) and (b). The speed of displacement of the peaks is calculated as ~$10^3$ m/s which corresponds to ion sound speed. Similar observations are also present in fig. 7 (b) $E_r$ plots. This observation indicates that the density gradient is associated with the generation of ion acoustic waves in plasma which is electrostatic in nature [21-22]. Hence, ambipolar electric field reversal and electrostatic heating is initiated during this time of plasma evolution.

E.    Verification of simulated results with experiment:

Figure 9(a) shows the line plot of simulated hot $T_e$ passes through the region very close to the discrete experimental data points. At 600 ns, the value of simulated hot $T_e$ is ~ 78 eV at 70 W of plasma absorbed power. At the same instant of time, the magnetron set power reaches 50 W and separately at steady state condition, the hot $T_e$ measured by Langmuir probe at this set power is ~ 72 eV.

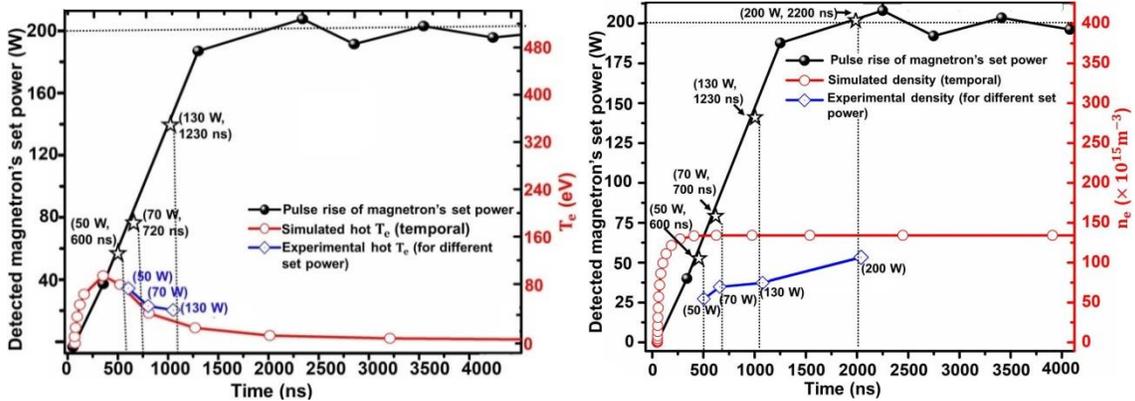

*Figure 9: (a) Simulated temporal hot $T_e$ variation along with experimental data, taken at discrete set power values. Plasma absorbed power during simulated hot $T_e$ evolution is fixed at 70 W. Experimental set power 50 W, 70 W and 130 W corresponds to plasma absorbed power 40 W, 50 W and 70 W respectively. Pulse rise of set power is also shown to verify the power levels during the ns to few μs periods. (b) Simulated temporal plasma density variation along with experimental results, obtained at the same set power values.*

Similarly, the measured hot $T_e$ at 70 W and 130 W are ~ 36 eV and 28 eV respectively which are very close to the simulated values. Whereas, in figure 9(b) the experimentally measured plasma density at these above three set power levels is 1.8-2 times less than the simulated values. But at 200 W set power, the measured plasma density (~$1.1 \times 10^{17} m^{-3}$) point lies near to the simulated value (~$1.3 \times 10^{17} m^{-3}$) because the plasma absorbed power (~70 W) at this set power is same as that of the simulated case.

## V.  Summary and conclusion

The paper describes spatial and temporal absorption of 2.45GHz microwave in a MDIS cavity volume for different low pressure regime. It also presented time evolution of plasma density and its effect of MW power deposition profile. Main power absorption phenomena is based on ECR mechanisms. In addition, other possible route for MW power absorption are off-resonance type which comprises of UHR heating and electrostatic Ion Acoustic wave heating. This off-ECR heating may dominate plasma density is above critical density (over dense condition) and MW fail to penetrate deep into the plasma. A detailed study of different power coupling mechanisms under over dense condition including polarization reversal, Doppler-shift and harmonic heating are reserved for future.